\documentclass[prb,aps,amssymb,showpacs,twocolumn]{revtex4}
\usepackage{amsmath}
\usepackage{amssymb}
\usepackage{amsthm}
\usepackage{amsfonts}
\usepackage{enumerate}
\usepackage{latexsym}
\usepackage{graphicx}

\newcommand{\beq}{\begin{equation}}
\newcommand{\eneq}{\end{equation}}

\input{epsf}

\begin{document}

\tolerance 10000

\newcommand{\vk}{{\bf k}}


\title{Maxwell Equation for the Coupled Spin-Charge Wave Propagation}
\author {B. Andrei Bernevig, Xiaowei Yu and Shou-Cheng Zhang}

\affiliation{Department of Physics, Stanford University,
         Stanford, California 94305 }

\begin{abstract}
We show that the dissipationless spin current in the ground state
of the Rashba model gives rise to a reactive coupling between the
spin and charge propagation, which is formally identical to the
coupling between the electric and the magnetic fields in the $2+1$
dimensional Maxwell equation. This analogy leads to a remarkable
prediction that a density packet can spontaneously split into two
counter propagation packets, each carrying the opposite spins. In
a certain parameter regime, the coupled spin and charge wave
propagates like a transverse ``photon". We propose both optical
and purely electronic experiments to detect this effect.
\end{abstract}

\pacs{72.25.-b, 72.10.-d, 72.15. Gd}

\maketitle

The generation and manipulation of spin current is essential to the
rapidly developing field of spintronics \cite{wolf2001}. For this
purpose, the analogy between photonics and spintronics greatly
helped our conceptual developments, and this analogy lead to the
celebrated Das-Datta proposal for a spin-field
transistor\cite{datta1990}. However, earlier attempts to realize
this conceptual device were based on the rather incomplete analogy
between the electron spin and the photon, and were plagued by many
issues such as low spin injection rate and the requirement of the
ballistic spin transport.

Recently, the remarkable phenomenon of the dissipationless spin
current has been theoretically predicted\cite{murakami2003}. A
electric field $E_k$ generates a spin current described by the
response equation
\begin{equation}
j_j^i = \sigma_s \epsilon_{ijk} E_k \label{spin_response}
\end{equation}
where $j_j^i$ is the current of the $i$-th component of the spin
along the direction $j$, $\epsilon_{ijk}$ is the totally
antisymmetric tensor in three dimensions and the spin Hall
conductivity $\sigma_s$ does not depend on impurities. Since both
the spin current and the electric field are even under the time
reversal, this equation describes a reactive response which does
not dissipate energy. One natural consequence of this equation is
the intrinsic spin Hall effect \cite{murakami2003, sinova2003},
which has been recently observed experimentally in the hole doped
systems \cite{wunderlich2004}. Another consequence is the
dissipationless spin current in the ground state
\cite{rashba2003}. In the above equation (\ref{spin_response}),
the electric field can be either externally applied, or can be
spontaneously generated in systems without inversion symmetry. In
a two dimensional electron gas (2DEG), the confining potential
along the $z$ direction breaks the inversion symmetry, and leads
to a internal electric field $E_z$ in the ground state. According
to equation (\ref{spin_response}), there is a spin current in the
ground state, $j_j^i = j_0 \epsilon_{ij}$, where $\epsilon_{ij}$
is the antisymmetric symbol in two dimensions with $i,j=x,y$.

In this work, we shall show that the dissipationless spin current in
the ground state makes the analogy between photonics and spintronics
formally exact. In $2+1$ dimensions, the electric field has two
components, while the magnetic field has only one component. If one
identifies them with the in-plane components of the spin density and
the charge density, respectively, the Boltzmann transport equation
for the coupled spin and charge wave is formally the same as the
Maxwell equation describing the electromagnetic fields, where the
``speed of light" is given by the Rashba coupling constant. This
behavior is in sharp contrast to the conventional Boltzmann equation
for the decoupled spin and charge dynamics in semiconductors, where
only purely diffusive, but no propagating motion is predicted
\cite{zutic2004}. The photonic analogy helps our understanding on
how density gradient and time dependence can generate spin density,
and leads to many novel predictions. We shall show that there is a
parameter regime, reachable experimentally, where the coupled
spin-charge wave propagates as a under-damped ``photonic mode". A
density packet will split spontaneously into two counter propagating
packets, each carrying the opposite spins. This mechanism enables
injection of spins and spin currents. The Boltzmann transport
equations for the Rashba model have been studied previously in the
diffusive region \cite{mishchenko2004,burkov2003}. The coupled
spin-charge wave propagation is a new result of this work.

A spin $1/2$ Hamiltonian which includes spin orbit coupling can be
written in the following general form:
\begin{equation}
H= \frac{p^2}{2m} + \lambda^i(p) \sigma^i, \;\;\; i=x,y,z
\end{equation}
\noindent where $\lambda^i(p)$ is a odd function of $p$, in order
to preserve the time reversal symmetry. This includes a wide range
of spin-orbit couplings, including 2D Rashba and Dresselhaus
couplings and the 3D spin splitting of the conduction band in
strained semiconductors \cite{bernevig2004}. The phase space
density distribution function $n_F(p,r,t)$ and the energy matrix
$\epsilon_F(p,r,t)$ are $2\times 2$ matrices, and can be
decomposed as:
\begin{eqnarray}
&&n_F(p,r,t)= n(p,r,t) + S^i(p,r,t) \sigma^i \nonumber \\
&&\epsilon_F(p,r,t) = \epsilon_s(p,r,t) + \epsilon_v^i(p,r,t)
\sigma^i, \;\;\; i=x,y,z \;\;\;\;\;\; \label{phasespacedensity}
\;\;\;
\end{eqnarray} \noindent In this letter we consider the system in
the absence of external fields, such that $\epsilon_F(p,r,t) =
p^2/2m + \lambda^i(p) \sigma^i$. The influence of electric and
magnetic fields on the system is described in a future longer
publication \cite{bernevigprb}. The Boltzmann equation reads:
\begin{widetext}
\begin{equation}
 \frac{\partial n_F(p,r,t)}{\partial t} -\frac{i}{\hbar}
[\epsilon_F(p,r,t), n_F(p,r,t)]  + \frac{1}{2} \{ \frac{\partial
\epsilon_F(p,r,t)}{\partial p_i}, \frac{\partial
n_F(p,r,t)}{\partial r_i}\}   -\frac{1}{2} \{ \frac{\partial
\epsilon_F(p,r,t)}{\partial r_i}, \frac{\partial
n_F(p,r,t)}{\partial p_i}\} = \frac{ n_F^{eq} - n_F(p,r,t)}{\tau}
\end{equation} \end{widetext} \noindent where the right hand side is
the collision term expressed in the relaxation time approximation,
and $n_F^{eq}$ is the equilibrium value of $n_F(p,r,t)$. $\tau$ is
the momentum relaxation time. Although this approximation does not
take into account the self-energy effects, it turns out to be
qualitatively and quantitatively correct, as we see from
comparison with the solution involving the self energy in some
special cases of spin-orbit coupling \cite{mishchenko2004,
burkov2003}. We trace out the matrix dependence of the
distribution function as well as that of the energy, and integrate
the continuity and the current equations over the Fermi volume
\cite{bernevigprb}. After linearization, we obtain
\cite{bernevigprb}:
\begin{equation}
 \frac{\partial n(r,t)}{\partial t} = D \frac{\partial^2
n(r,t)}{\partial r_i^2} - \frac{\partial \lambda^l(p)}{\partial p_i}
\frac{\partial S^l(r,t)}{\partial r_i}
\end{equation} \noindent
\begin{widetext}
\begin{equation}
\frac{\partial S^k(r,t)}{\partial t} = D \frac{\partial^2
S^k(r,t)}{\partial r_i^2} - \frac{\partial \lambda^k(p)}{\partial
p_i} \frac{\partial n(r,t)}{\partial r_i}  + \frac{4mD}{\hbar}
\epsilon_{ijk} \frac{\partial \lambda^i}{\partial p_r}
\frac{\partial S^j(r,t)}{\partial r_r}  - \left( \frac{2m}{\hbar}
\right)^2 D \left(\frac{\partial \lambda^i}{\partial p_r}
\frac{\partial \lambda^i}{\partial p_r} S^k (r,t)  - \frac{\partial
\lambda^i}{\partial p_r} \frac{\partial \lambda^k}{\partial p_r}
S^i(r,t) \right) \label{generalcontinuityequations} \;\;\;
\end{equation} \end{widetext} \noindent
where $D = \frac{\langle p_F^2\rangle \tau} {2 m^2}$ is the
diffusion constant and $\mu =\frac{e \tau}{m}$ is the mobility, and
the carriers are electrons of charge $-e$. Aside from the self
energy renormalizations, Eq.[\ref{generalcontinuityequations}] gives
the same result as \cite{mishchenko2004,burkov2003} when
particularized to the Rashba-spin-orbit coupling. The last term in
the spin continuity equation represents the spin relaxation due to
Dyakonov-Perel (DP) mechanism \cite{dyakonov1971}. In the case of
Rashba systems, the spin orbit coupling is $\lambda^{i} = \alpha
\epsilon_{ijz} p_j$, $\alpha$ has the dimension of velocity, and the
continuity equations become (from now on we use $\partial_i =
\partial / \partial r_i$ and $\partial_t = \partial /\partial t$) :
\begin{eqnarray}
&& \partial_t n  = D\partial_i^2 n - \alpha
\epsilon^{liz} \partial_i S^l  \nonumber \\
&& \partial_t S^k = D\partial_i^2 S^k  - \alpha \epsilon^{kiz}
\partial_i n +  \sqrt{\frac{D}{\tau_s}}(\delta_{kz}
\partial_i
S^i - \partial_k S^z ) - \nonumber \\
&& - \frac{1}{\tau_s}(S^k \mathbf{+} \delta_{kz} S^z) \;\;\;
\label{RashbaBoltzmann}
\end{eqnarray} \noindent where $r_i =(x,y)$ ($i=1,2$) since charge
and spin motion is now confined entirely to the 2D plane. Within
the current microscopic approximation, the DP spin relaxation time
is given by $\tau_s^{-1}=(\frac{2m\alpha}{\hbar})^2D$, however, in
the subsequent discussions, we shall treat $D$ and $\tau_s$ as
independent, phenomenological parameters.

Let $S^\mu = (S^k, S^z)$, $\mu= x,y,z, \;\; k=x,y$. We can write the
two dimensional vector $S^k(r,t), \; k=x,y$ in the most general form
as a sum of a longitudinal vector $S^k_L(r,t)$ and a transversal
vector $S^k_T(r,t)$:
\begin{equation}
S^k = S^k_L  + S^k_T; \;\;\; \partial_k S^k_T = 0, \;\;\;
\epsilon_{ij} \partial_i S^j_L = 0
\end{equation}
\noindent Substituting this decomposition into Eq.
(\ref{RashbaBoltzmann}), we find two sets of coupled equations:
\begin{eqnarray}
&& \partial_t n = D \partial_i^2 n  - \alpha \epsilon_{ki}
\partial_i S^k_T
 \nonumber \\
&& \partial_t S^k_T = D \partial_i^2 S^k_T - \alpha \epsilon_{ki}
\partial_i
n - \frac{1}{\tau_s} S^k_T \nonumber \\
&& \partial_k S^k_T = 0 \;\;\; \label{Maxwell}
\end{eqnarray} and
\begin{eqnarray}
&& \partial_t S^k_L = D \partial_i^2 S^k_L  -
\sqrt{\frac{D}{\tau_s}} \partial_k S^z - \frac{1}{\tau_s} S^k_L
\nonumber \\
&& \partial_t S^z = D \partial_i^2 S^z + \sqrt{\frac{D}{\tau_s}}
\partial_k
S^k_L - 2 \frac{1}{\tau_s} S^z \nonumber \\
&& \epsilon_{ij} \partial_i S^j_L = 0\;\;\;
\end{eqnarray} \noindent
Hence the charge density couples only to the transverse spin
component, while $S^z$ couples only to the longitudinal spin
component in a purely diffusive fashion\cite{burkov2003}. In the
spin continuity equation of Eq. (\ref{Maxwell}), we see that the
$\alpha$ term is nothing but the divergence of the dissipationless
spin current in the ground state $j_i^k=\alpha\epsilon_{ki} n$. We
shall see that this term plays the crucial role leading to the
coupled spin-charge propagation.

At this point we come to a remarkable realization that the
Boltzmann equation (\ref{Maxwell}) for the coupled charge and
transverse spin transport is exactly the Maxwell's equation in
$2+1$ dimensions! In order to facilitate the comparison, let us
first focus on the large spin-orbit coupling limit, where we
neglect the $D$ and the $\frac{1}{\tau_s}$ terms in Eq.
(\ref{Maxwell}). In $2+1$ dimensions, the source-free Maxwell
equations are given by
\begin{eqnarray}
&& \partial_\nu F_{\mu\nu}=j_\mu=0 \label{source}\\
&& \epsilon^{\mu\nu\rho}\partial_\mu
F_{\nu\rho}=0\label{source-free}\;\;\;
\end{eqnarray} \noindent
where $\mu=(0,x,y)$. In $2+1$ dimensions, the magnetic field has
only one component, given by $B_z=F_{xy}$, and the two components
of the electric field are given by $E_i=F_{0i}$. If we make the
identification $n \rightarrow B_z$, $S^i_T \rightarrow E_i$ and
$\alpha \rightarrow c$, we see that the three Boltzmann transport
equations in Eq. (\ref{Maxwell}) are exactly the three Maxwell's
equations in the vacuum of $2+1$ dimensions, namely the Faraday's
law of induction, the Ampere-Maxwell law, and the Gauss's law.
More generally, the $\frac{1}{\tau_s} S^k_T$ term can be
interpreted as light propagation in a metallic media, with the
current density in the Ampere's law given by the Ohm's law, and
the $D$ terms can be interpreted as due to light diffusion in a
random media.

\begin{figure}[h]
\includegraphics[scale=0.34]{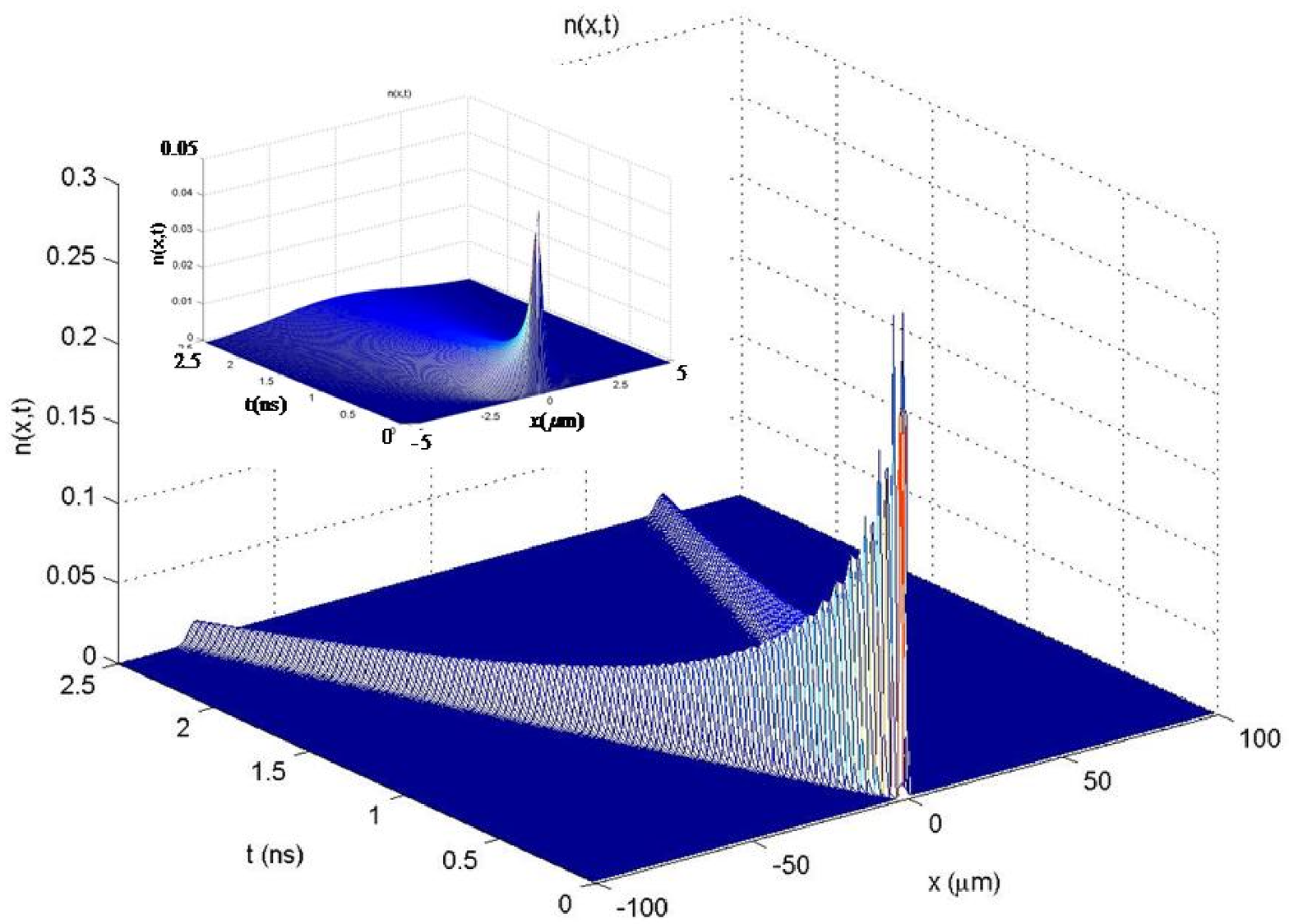}
\includegraphics[scale=0.34]{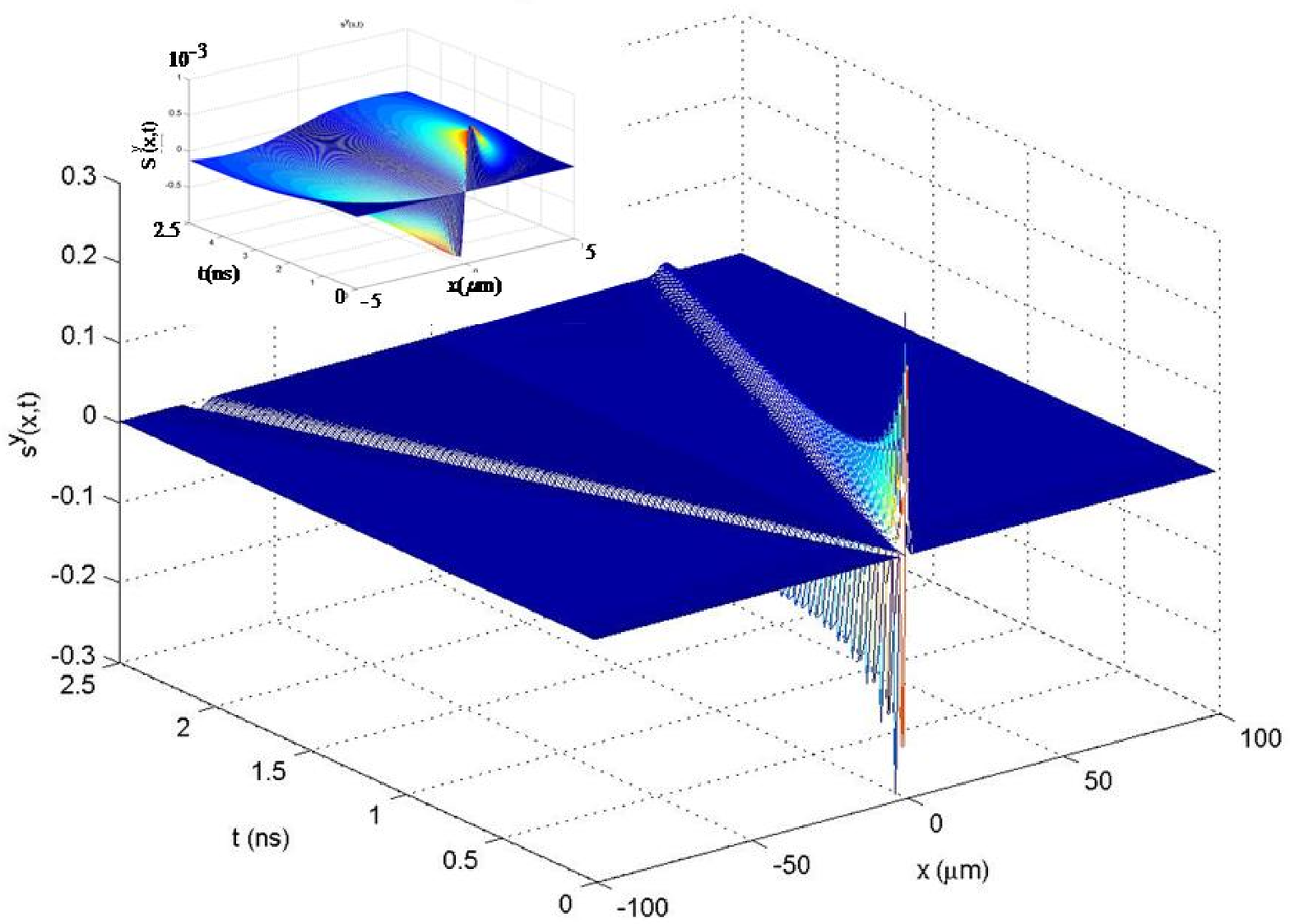}
\caption{ Charge and spin density for $\tau_s =1ns$, $\alpha =
3\times 10^4 m/s$ and $D=10^{-3}m^2/s$. We see propagation
  over distances of more than $100 \mu m$. \textbf{Inset:} Charge
  and spin density in the diffusive
  regime, for small values of $\alpha$ ( $\tau_s =1ns$, $\alpha =
 10^2 m/s$ and $D=10^{-3}m^2/s$)  has the typical Gaussian
  decay.} \label{chargedensity}
\end{figure}

In conventional theories without spin-orbit coupling, the electron
transport semiconductors is purely diffusive. However, we see that
in the limit of strong spin-orbit coupling, there is a regime
where a propagating, coupled spin-charge wave mode is possible. If
we neglect the diffusion and the lifetime terms for the time
being, we find that the most general solution to the initial
condition of $n(x,y,t=0) = f(x)$ and $S^i_T(x,y,t=0)=0$ is given
by:
\begin{eqnarray}
&& n(x,t) = \frac{1}{2}(f(x + \alpha t) + f(x- \alpha t)) \nonumber \\
&&  S_T^y(x,t) = \frac{1}{2}( f(x+ \alpha t) - f(x - \alpha t) )
\end{eqnarray}
\noindent We see that an initial density wave packet spontaneously
splits into two counter-propagating packets, each carrying the
opposite spin. This phenomenon can be elegantly interpreted in the
``photonic" language. In $2+1$ dimensions, the magnetic field is
always pointing along the $+\hat{z}$ direction. Since the
propagation vector ${\bf k}$, being proportional to the Poynting
vector, is given by ${\bf k}\propto {\bf E}\times  {\bf B}$, it
uniquely determines the direction of the transverse electric field.
Translating from the ``photonic" language into the ``spintronic"
language, we see that the mode propagating along the $+\hat{x}$ has
spins along the $+\hat{y}$ direction, while the mode propagating
along the $-\hat{x}$ has spins along the $-\hat{y}$ direction. The
split wave packets carries a spin current $J_x^y$, which is a
reflection of the spin current in the ground state of the Rashba
model. For a simple estimate, $\alpha = 3 \times 10^4 m/s$, and
hence the mode will cross a sample of $1 \mu m$ length in $30 ps$.
Considering that the spin coherence time in these samples can be
larger than $1 ns$, it means that the propagation time over $1 \mu
m$ distance is well shorter than $30$-th part of the spin relaxation
time and can hence be very useful for spin manipulation.

We now consider the more general situation including diffusion and
relaxation. We suppose that a one dimensional stripe of charge
density has been created, say by transient grating
\cite{gedik2003}. The initial density is given by
$n(x,y;t=0)=\delta(x)$. The solution to the full equations give
$S^z(x,t) =S^y(x,t) =0$ while $S^y(x,t)$ is generated by the
spin-orbit coupling:
\begin{equation}
n(x,t) = \int \int \frac{1}{(2 \pi)^2} \frac{i\omega + D q^2
+\frac{1}{\tau_s} }{- (\omega - \omega_1) (\omega - \omega_2)} e^{i
(\omega t - q x)} d \omega dq
\end{equation}
\begin{equation}
S^y(x,t) = \int \int \frac{1}{(2 \pi)^2} \frac{i \alpha q }{ (\omega
- \omega_1) (\omega - \omega_2)} e^{i (\omega t - q x)} d \omega dq
\end{equation}
\noindent where $\omega_1, \omega_2$ are the characteristic
frequencies of the system:
\begin{equation}
\omega_{1,2} = i(D q^2 + \frac{1}{2\tau_s})  \pm \sqrt{\alpha^2 q^2
- \frac{1}{(2 \tau_s)^2}}
\end{equation}
\noindent We recognize the propagating mode inside the square
root. For momenta $q> 1/2\tau_s \alpha$ both characteristic
frequencies contain real parts and hence describe propagating
waves. However, $q$ must not be as large as to cause damping due
to the term $ Dq^2 t$. The condition for this gaussian damping to
be small is $ D q^2 \tau_s <1$ for $q \sim 1/ \tau_s \alpha$.
Therefore, the condition for the regime where a propagating mode
could exist is then given by:
\begin{equation}
\alpha > \sqrt{\frac{D}{\tau_s}}
\end{equation}
\noindent This condition can be satisfied in samples where $\alpha
=3 \times 10^4 m/s$ and $D=10^{-3} m^2/s$,  with $\tau_s$ longer
than $1 ns$ \cite{kato2004}. In this case,
$\sqrt{\frac{D}{\tau_s}}=10^3$, much smaller than $\alpha$.

\begin{figure}[h]
\includegraphics[scale=0.44]{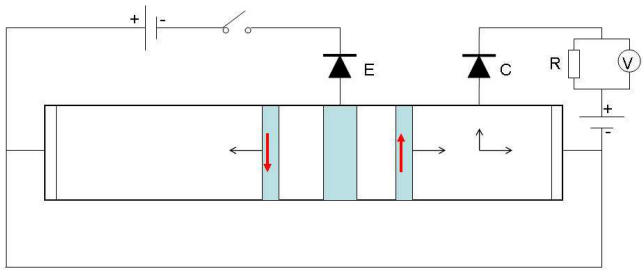}
\caption{A modified version of the classic Haynes-Shockley
experiment. A density packet injected by the emitter spontaneously
splits into two counter propagating packets with opposite spin.
Unlike the settings of the Haynes-Shockley experiment, one of the
two packets propagates to the collector without experiencing a
sweeping electric field. The time delay between the injection
pulse the the collecting pulse gives a purely electric
determination of the Rashba spin-orbit coupling constant.}
\label{shockley}
\end{figure}

In the limit of very long $\tau_s \rightarrow \infty$ the
integrals can be solved exactly. We give the expression of $S^y$
in this limit:
\begin{equation}
S^y(x,t) = \frac{1}{4 \sqrt{\pi}}\frac{1}{\sqrt{Dt}} \left[
e^{-\frac{(\alpha t -x)^2}{4Dt}} -e^{-\frac{(\alpha t +x)^2}{4Dt}}
\right]
\end{equation}
\noindent The propagating mode is $\alpha t \pm x =0$ where either
one of the damping gaussian exponentials becomes unity. The spin
symmetry is odd in $x$, the spins propagating in the positive and
negative $x$ axis directions having opposite polarization. Note
that for diminishing spin-orbit coupling $\alpha \rightarrow 0$
the spin density also vanishes, as it should. For finite $\tau_s$
in a stationary phase-type approximation the spin-density solution
above gets multiplied by an exponential factor $\exp(-t /2
\tau_s)$. Impressively, both spin and charge can propagate over
distances well in excess of $100 \mu m$ and for times well in
excess of $10 \tau_s$, (the full time scale is not plotted in
Fig[\ref{chargedensity}]).

We now propose several experiments to test the coupled spin-charge
wave predicted in this work. One could inject the density packet
optically, and detect the splitting of the density packet and the
associated spin orientation by optical Kerr rotation. One could also
detect the spin orientation through the circularly polarized
luminance from the recombination with the majority carriers.
Alternatively, one could detect the propagation of the density
packet purely electrically, by a modified version of the classic
Haynes-Shockley experiment\cite{haynes1951}. Fig. (\ref{shockley})
describes a narrow sample with light p-doping. Two rectifying
metal-to-semiconductor point contacts are forward and reverse
biased, respectively, to serve as emitter and collector electrodes.
After turning on the emitter pulse, a electron density packet is
injected into the sample. In conventional Haynes-Shockley setup, the
electron packet would be swept to the collector electrode by a
electric field. In our case, no sweeping electric field is applied,
but the density packet will spontaneously split into two counter
propagating packets with the opposite spin orientation, with a
velocity directly given by the Rashba coupling constant $\alpha$.
When the right moving packet is captured by the collector electrode,
a voltage pulse is registered. From the time-delay and the shape of
the voltage pulse, one can determine the Rashba coupling constant
and the diffusion constant by purely electric means. This experiment
illustrates the fact that the injected density pulse can take
advantage of the spin current in the ground state, and propagate
without any applied voltage.

\section{Acknowledgements}
The authors want to thank D. Awschalom, H.D. Chen, Y. Kato, J.
Orenstein and J. Zaanen for useful discussions. B.A.B. acknowledges
support from the Stanford Graduate Fellowship Program. This work is
supported by the NSF under grant numbers DMR-0342832 and the US
Department of Energy, Office of Basic Energy Sciences under contract
DE-AC03-76SF00515.

\end{document}